# Room-Temperature Exciton-Polariton Condensation in a Tunable Zero-Dimensional Microcavity

*Fabio Scafirimuto,[1] Darius Urbonas,[1] Ullrich Scherf,[2] Rainer F. Mahrt[1], and Thilo Stöferle[1*]*

[1] IBM Research – Zurich, Säumerstrasse 4, 8803 Rüschlikon, Switzerland

[2] Macromolecular Chemistry Group and Institute for Polymer Technology, Bergische Universität Wuppertal, Gauss-Strasse 20, 42119 Wuppertal, Germany

## ABSTRACT

We create exciton-polaritons in a zero-dimensional (0D) microcavity filled with organic ladder-type conjugated polymer in the strong light-matter interaction regime. Photonic confinement at wavelength scale is realized in the longitudinal direction by two dielectric Bragg mirrors and laterally by a sub-micron Gaussian-shaped defect. The cavity is separated into two parts allowing nanometer position control and enabling tuning of exciton and photon fractions of the polariton wavefunction. Polariton condensation is achieved with non-resonant, picosecond optical excitation at ambient conditions and evidenced by a threshold behavior with non-linear increase of the emission intensity, line narrowing and blue-shift of the emission peak. Furthermore, angular emission spectra show that condensation occurs in the ground state of the 0D-cavity, and first order coherence measurements reveal the coherent nature. These experiments open the door for polariton quantum fluids in complex external potentials at room temperature.

# MANUSCRIPT

Exciton-polaritons are light-matter excitations generated by the coupling of photons inside a high quality factor optical resonator with an embedded optically active material. By matching the energy of the optical resonance (cavity photons) and the material's opto-electronic resonance transition (excitons, i.e. bound electron-hole pairs) it is possible to observe the characteristic anti-crossing behavior that reflects the emergence of new eigenstates.[1] These exciton-polaritons are quasi-particles of bosonic nature. At sufficiently high density, they can undergo non-equilibrium Bose-Einstein condensation and form a polariton condensate.[2,3] This macroscopic quantum phenomenon has been observed in various inorganic semiconductors[4] and more recently in organic materials,[5,6] even up to room-temperature, which is very appealing in view of future possible technological applications.[7,8]

Motivated by the success of ultracold atomic gases,[9] the use of a condensate for analog quantum simulations of solid state and non-equilibrium toy-model Hamiltonians[10] received enormous attention, recently. Here it is crucial to be able to create arbitrary potential landscapes for the quantum fluid to emulate different kinds of crystal lattices and the intricate effects that arise from the various band structures, e.g. the Dirac cones in graphene. Furthermore, through strong lateral confinement, lower dimensional geometries like one-dimensional nanowires can be simulated. Lithographic definition of small mesas,[11] deep etching of the whole microcavity structure[12] and patterning of thin metal layers[13] have been used to realize the confinement for semiconductor microcavities. Wavelength tunability can be achieved through separating the two cavity mirrors and mounting them on nanopositioners.[14] Gaussian-shaped nanoscale defect structures are enabling much tighter, wavelength-scale lateral confinement[15] with theoretical cavity quality factors above $10^5$ that can be fabricated using focused ion beam milling.[16]

In this work, we report the observation of room temperature polariton condensation in a tunable Gaussian shaped microcavity based on an organic polymer. The material used is a methyl-substituted ladder-type poly(p-phenylene) (MeLPPP),[17] which is a π-conjugated organic semiconductor polymer with a comparably rigid backbone, that was already used to show polariton condensation in flat Fabry-Pérot cavities.[5] We build on a configuration described by Urbonas *et al.*[18] enabling tunable room temperature strong coupling. It consists of two cavity halves that are mounted on XYZ nanopositioners (Figure 1a, see Methods for details): One part consists of a glass mesa with a Gaussian defect and a subsequently deposited distributed Bragg reflector (DBR). The other one consists of a DBR and spin-coated thin film of MeLPPP, protected by a thin inorganic layer.

In addition to forming a 0D-localization through the Gaussian defect, this configuration provides control over the distance between the mirrors, and hence, allows to vary the composition of the polariton wavefunction between more photon-like and more exciton-like. To observe the strong coupling regime manifested by the splitting of the cavity mode into upper and lower polariton branches, we illuminate the sample with a broadband halogen lamp and record the transmission spectrum while tuning the cavity resonance across the exciton energy (Figure 1b). In addition to the lowest Laguerre-Gaussian mode (LG00) from the cylindrically symmetric 0D-cavity, we also observe the planar cavity mode (PC) because our probe beam is larger than the Gaussian defect. By fitting the upper and the lower polariton branches of the LG00 mode with a coupled oscillator model, we extract a Rabi splitting of $2\Omega$ = 60 meV. The change of the wavefunction composition when tuning the cavity resonance can be inferred from calculating the Hopfield coefficients (Figure 1c).

To achieve polariton condensation, we excite the microcavity off-resonantly at 3.1 eV photon energy under normal incidence, outside the DBR bandgap. We use a single-mode photonic crystal fiber to obtain a small excitation spot close to the dimensions of the Gaussian defect and to stretch the excitation laser pulse to ~12 ps duration. The emitted light is either detected by a camera (real space image), or a spectrograph equipped with a two-dimensional detector or is feed into a Michelson interferometer (see Methods for details). Above a certain threshold we find non-linear behavior in the total emitted intensity, a narrowing of the linewidth of the polariton mode and a blue-shift of the peak position. Comparing the angular emission pattern below and above threshold, a dramatic increase of the polariton population in the lowest energy mode is observed. Finally, first order coherence measurements allow for estimating the temporal coherence length.

**RESULTS**

The 0D-character of the microcavity has a pronounced impact on the angular emission spectrum. While the planar cavity (PC) has a parabolic dispersion relation, the Laguerre-Gaussian modes

LG*nl* have discrete energies, where *n* and *l* give the number of nodes in the radial and azimuthal direction, respectively. At low excitation power (Figure 2a), we observe the lowest azimuthal order LG00 and the first order LG01 from the Gaussian defect and the planar cavity mode (PC) from the area pumped in the vicinity of the Gaussian defect. At high excitation power (Figure 2b), the situation changes completely: The polaritons condense in the LG00 mode (lowest energy mode) while the other modes are so weakly populated that they are barely observable. Already from this measurement the non-linear increase of the emission, the line narrowing and the blue-shift, all indications for polariton condensation, are visible. Furthermore, it shows that indeed condensation in the lowest energy state is occurring. Polariton lasing[12] from the higher LG01 mode has been observed only occasionally when the cavity was detuned to an extremely photonic polariton composition (>97% photon contribution).

To obtain quantitatively the threshold excitation density $P_{th}$ for the condensation we measured the angularly and spatially integrated emission spectrum versus the excitation fluence. The recorded spectra of the lowest polariton branch of the LG00 mode were fitted with a Gaussian peak allowing to extract its area, width and center. The low emission intensity far below the threshold limits the range where the fit works reliably to as low as ~0.5 $P_{th}$. In Figure 3a, the total emitted intensity (area of the peak) versus excitation power displays the characteristic non-linear behavior due to the stimulated scattering into the polariton ground state. Furthermore, the peak width exhibits a significant drop of the linewidth, corresponding to transition towards the condensate regime. For some samples, we find peak widths down to 300 μeV, which is almost one order of magnitude narrower than for planar cavities,[5] and corresponds to the Fourier limit imposed by the pump duration. This points towards different condensation dynamics in the 0D-cavity, where only one discrete lowest state exists, and which is essentially free of photonic spatial disorder and potential influence of instabilities of the polariton reservoir.[19] The peak center energy as displayed in Figure 3c shows the typical blue-shift which is due to the interaction among the polaritons and the exciton reservoir[20] and saturation effects.[19] In the range of excitation powers and polariton energies studied, the behavior is well-described by a linear to sub-linear increase of energetic shift with excitation power (power-law with exponent between 0.3-1.0) and does not exhibit a pronounced jump when the condensate regime is entered. It must be noted that the variability of these quantities for different spots of the conjugated polymer can be quite high, e.g. the threshold fluence can vary up to 50% at a particular energy.

Finally, we study the first order coherence of the condensate by coupling the polariton emission into a Michelson interferometer. The emission is split to two interferometer arms, where one can be delayed and is spatially inverted with respect to the other, and then both real space images are focused on a camera. As the light originates from the Gaussian-shaped LG00 mode, the interferogram is a Gaussian with superimposed fringes (Figure 4a), and the fringe period corresponds to the relative angle under which the beams where overlapped on the camera. We obtain the temporal coherence from the fringe amplitude as a function of interferometer arm delay by Fourier transformation of the images (Figure 4b). Below threshold, the polariton emission has a very short coherence time, and we find a very short decay of the coherence within 10 fs. Above threshold, the polariton condensate gives rise to dramatically prolonged coherence by about two orders of magnitude, up to a few picoseconds. For some spatial positions on the polymer, the coherence above threshold shows exponential and for others Gaussian fall-off characteristics,[21] which can point towards different magnitude of fluctuations dependent on the local excitonic disorder.[22]

## **CONCLUSIONS**

We demonstrated exciton-polariton condensation in a 0D-microcavity at ambient conditions, as evidenced by a comprehensive set of experimental results. The sub-micron sized tunable defect structure establishes a promising platform for the investigation of quantum fluids in arbitrary potential landscapes for analog quantum simulations.

## **METHODS**

**Sample Preparation**. To realize the top part of the cavity we firstly define a mesa structure (~200 μm diameter and ~30 μm height) by back-etching on a borosilicate substrate, then we pattern the Gaussian defects (depth ~30 nm and full width half maximum (FWHM) ~1000 nm) by focused ion beam milling. Finally, we deposit 6.5 double layers of $Ta_2O_5/SiO_2$ forming the DBR by magnetron sputtering. The bottom part is realized by sputtering 9.5 double layers on a borosilicate substrate. Subsequently, we deposit a 35 nm-thin layer of the MeLPPP polymer by spin coating from a toluene solution (1% concentration) and complete the structure by sputtering a protective layer of 10 nm $SiO_2$ on top.

**Optical Characterization**. The two cavity parts are mounted on independent XYZ stages which allow a nanometer control of the positions. The microcavity is excited either with a multimode fiber-coupled halogen lamp for the transmission measurements or with a single mode fiber-coupled (2 m polarization maintaining photonic crystal fiber, 5 µm mode diameter) frequency-doubled regenerative amplifier seeded by a mode-locked Ti:sapphire laser for all other measurements . The initial pump pulse is 100 – 200 fs with a repetition rate of 1 kHz. The pump intensity at the fiber output is controlled with a movable gradient filter. The excitation beam is focused through the top cavity part with a long working distance apochromatic microscope objective (100X, numerical aperture NA = 0.5) to a spot with FWHM = 2 – 3 µm on the sample. The emitted light from the cavity is collected through the bottom part with an apochromatic microscope objective (20X, NA = 0.5) which is mounted on a third XYZ nanopositiong stage. Suitable long-pass filters are used to block the excitation light. The emission is detected either by an imaging spectrograph (liquid $N_2$-cooled CCD (charge-coupled device), 0.5 m focal length, 1800 lines/mm grating for the spectra and 300 lines/mm for the dispersion curves) or by a cooled CCD camera (for the real-space images). For the dispersion curves, the spectrograph performs *k*-space imaging with an entrance slit width of 50 µm. For the coherence measurements, the emitted light is split by a non-polarizing beam-splitter cube and recombined and focused after the Michelson interferometer on a cooled CCD camera. In one interferometer arm, a hollow retroreflector (used to invert the image) is mounted, and a motorized linear stage and an additional piezo provide the variable time delay.

**AUTHOR INFORMATION**

**Corresponding Author**

* tof@zurich.ibm.com (T.S.)

**Author Contributions**

The manuscript was written through contributions of all authors. All authors have given approval to the final version of the manuscript.

**Notes**

The authors declare no competing financial interest.

**ACKNOWLEDGMENTS**

We are grateful to W. Riess, G.-L. Bona and A. Imamoglu for discussions, U. Drechsler, R. Stutz, S. Reidt and M. Sousa for help with the sample preparation and characterization. This work was partly supported by the Swiss State Secretariat for Education, Research and Innovation (SERI) and the European Union's Horizon-2020 framework programme through the Marie-Sklodowska Curie ITN networks PHONSI (H2020-MSCA-ITN-642656) and SYNCHRONICS (H2020-MSCA-ITN-643238).

**TOC Graphics**

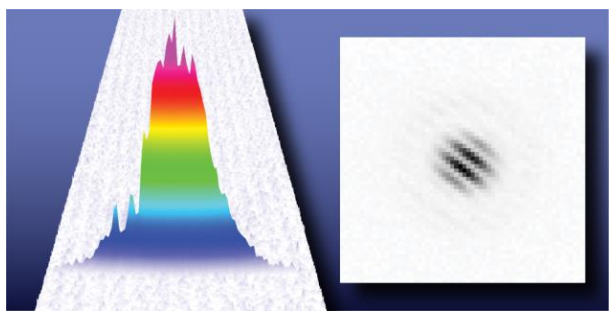

**FIGURES**

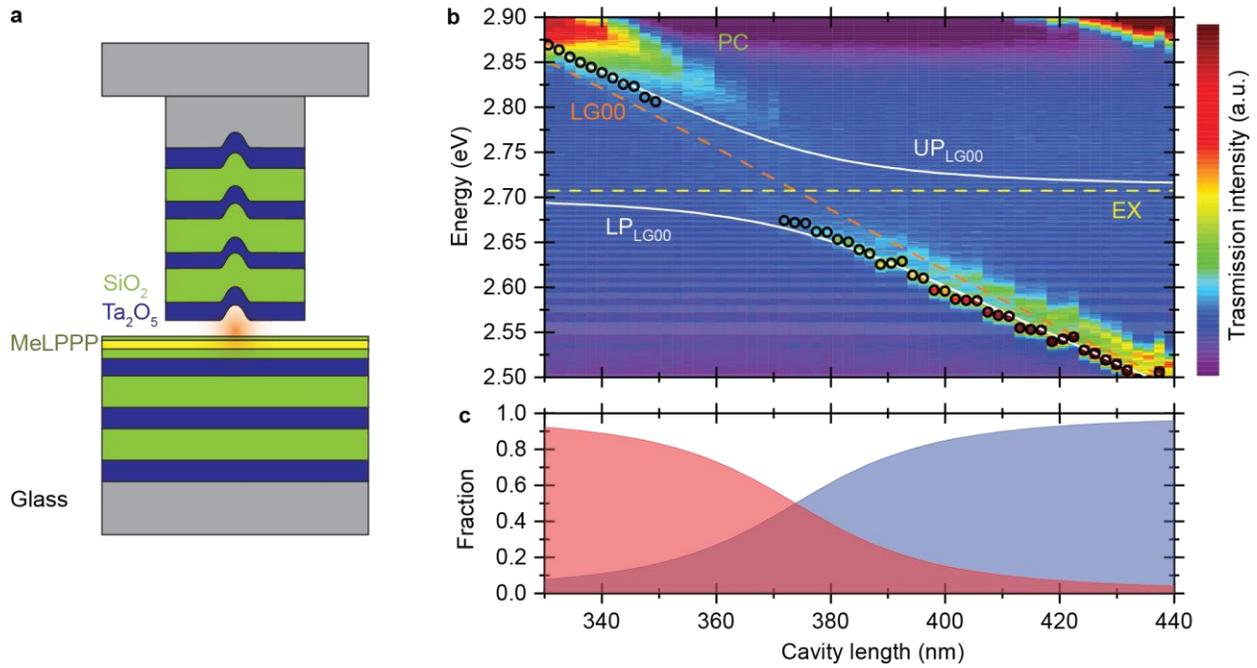

**Figure 1.** (a) Sketch of the tunable Gaussian defect cavity. (b) Transmission spectrum as a function of distance between the cavity halves, showing the avoided crossing of the localized mode from the Gaussian defect (LG00, orange dashed line) and the exciton of the MeLPPP (EX, yellow dashed line) into lower (LP$_{LG00}$) and upper (UP$_{LG00}$) polariton branches. The black circles are extracted transmission peak centers from the measurement whereas the solid white lines are fits to these data points with a coupled oscillator model. (c) Composition of the polariton wavefunction of the LP$_{LG00}$ branch (photonic = blue, excitonic = red) calculated from the Hopfield coefficients when tuning the cavity length.

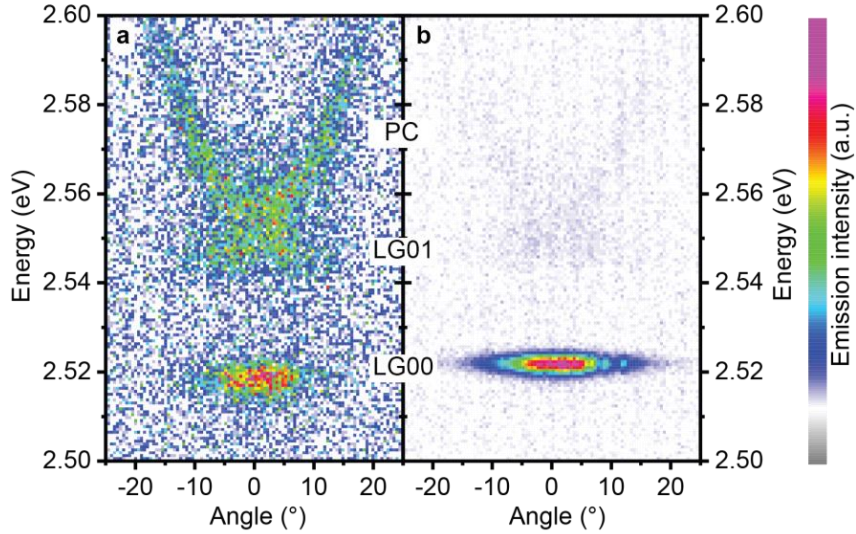

**Figure 2.** Angular dispersion of the emitted light measured with an imaging spectrograph in quasi-momentum-space configuration. The color scale is normalized to the intensity maximum in each plot, and the exposure times were 60 s for the left and 10 s for the right graph. (a) At an excitation fluence below threshold (0.5 $P_{th}$), the parabolic dispersion of the planar cavity (PC) together with flat bands of the LG00 and the higher order LG01 are observed. (b) Above threshold (2 $P_{th}$), the mode-structure is similar but the intensity distribution is completely changed. While the LG01 and the PC modes are barely visible, the lowest mode LG00 is strongly increased because of the stimulated scattering into the condensate. It is notable that all modes exhibit a slight blue-shift due the polariton-exciton interaction and saturation effects.

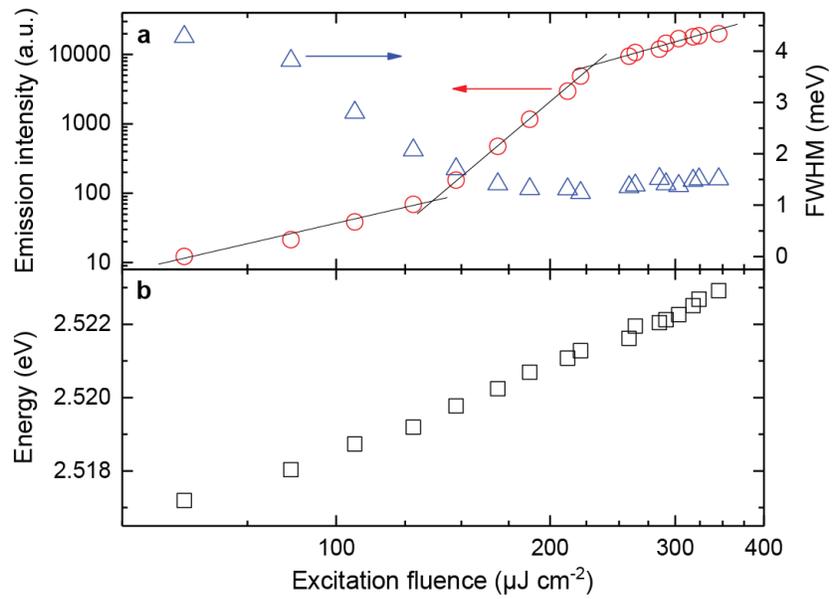

**Figure 3.** (a) Emission intensity versus excitation fluence (red open circles, left axis, log-log scale) together with emission linewidth (blue open triangles, right axis). A threshold for the super-linear increase is observed at around 130 µJ cm$^{-2}$ (lines are guide to the eye). This coincides with a drop of the linewidth from ~4.5 meV below threshold to ~1.5 meV above threshold. (b) Power dependent blue-shift of the polariton emission.

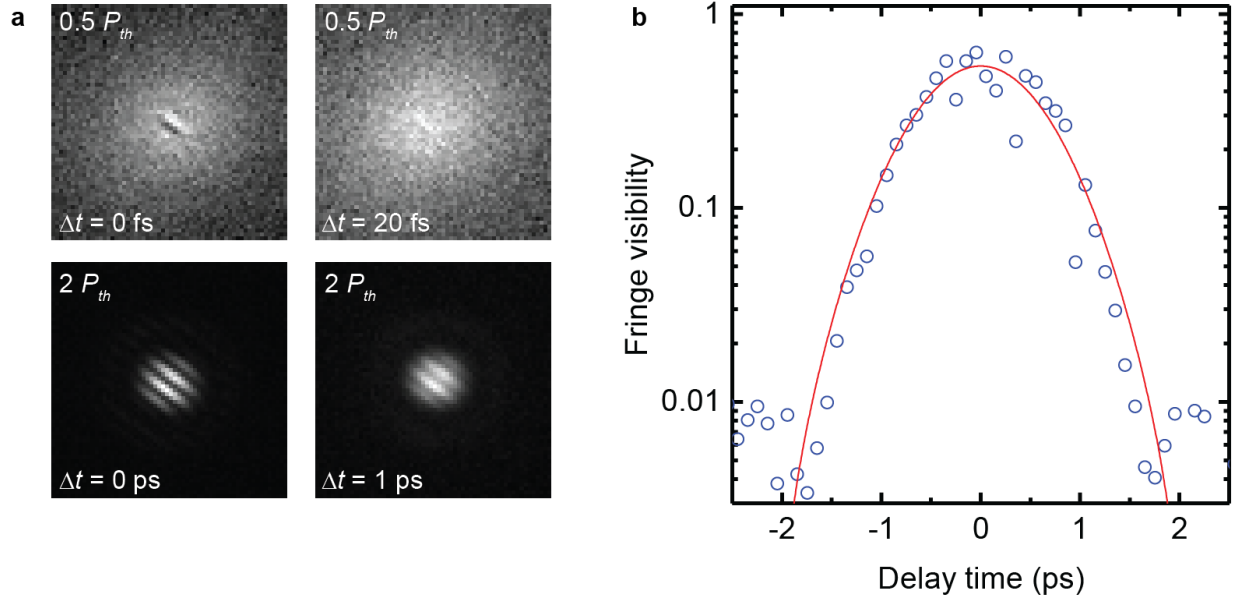

**Figure 4.** Fringe visibility of the polariton emission as a function of relative delay time $\Delta t$ between the Michelson interferometer arms. (a) Exemplary recorded real space images (dimensions 6.6 × 6.6 μm, gray scale normalized for each image). Below threshold, the coherence lasts only about 10 fs. Above threshold, the polariton condensate is formed, and the coherence is extended by two orders of magnitude up to more than a picosecond. The imaging exposure times are 30 s for the dataset below threshold and 1 s above threshold. (b) For a quantitative analysis of the condensate temporal coherence, the fringe visibility is extracted from the Fourier transformation of the images by using the magnitude at the fringes' spatial period. The solid line shows a Gaussian fit to the data. The fringe visibility is normalized to the value deduced directly from the raw images at zero time delay, giving maximum fringe amplitudes of 60-70% for both datasets below and above threshold.